\documentclass[]{IEEEtran}
\usepackage{wrapfig}
\usepackage{amsfonts}
\usepackage{amssymb,epic,eepic}
\usepackage{amsmath}
\usepackage{dcolumn}
\usepackage{bm}
\usepackage{bbm}
\usepackage{verbatim}
\usepackage{color}
\usepackage{amsthm}
\usepackage{tikz}
\usepackage{graphicx}
\usetikzlibrary{arrows,positioning,decorations.pathmorphing,decorations.markings,shapes,fadings}

\newcommand{\nn}{{\mathbb N}}

\newcommand{\eps}{{\varepsilon}}        

\newcommand{\bA}{\mathbf A}
\newcommand{\bB}{\mathbf B}
\newcommand{\bC}{\mathbf C}
\newcommand{\eins}{{\mathbbm{1}}}

\newtheorem{theorem}{Theorem}

\newtheorem{conjecture}{Conjecture}

\newtheorem{definition}{Definition}
\newtheorem{example}{Example}

\newtheorem{lemma}{Lemma}

\newtheorem{remark}{Remark}

\newcommand{\tr}{\mathrm{tr}}
\newcommand{\supp}{\mathrm{supp}}

\DeclareMathOperator{\linspan}{span}
\DeclareMathOperator{\ran}{ran}

\begin{document}
\title{Deducing Truth from Correlation}
\author{J.~N\"otzel, W.~Swetly
\thanks{
J.~N\"otzel had a joint affiliation with the Lehrstuhl f\"ur Theoretische Informationstechnik at Technische Universit\"at M\"unchen in 80290 M\"unchen and F\'{\i}sica Te\`{o}rica: Informaci\'{o} i Fen\`{o}mens Qu\`{a}ntics, Universitat Aut\`{o}noma de Barcelona in ES-08193 Bellaterra (Barcelona), Spain. He is now with the Technische Universit\"at Dresden, Communications Laboratory, 01069 Dresden, Germany.}
\thanks{
W. Swetly is affiliated with the Munich Center for Mathematical Philosophy, Ludwig-Maximilians-Universitä\"at, 80539 M\"unchen, Germany and has been with the Lehrstuhl f\"ur Datenverarbeitung, Technische Universit\"at M\"unchen, 80290 M\"unchen, Germany during the preparation of the first draft of this work.}
\thanks{J.N. thanks Holger Boche for continuously supporting his research efforts, and W.S. thanks Stephan Hartmann and Klaus Diepold for the great opportunities from 2012 to 2014. He is especially grateful to his academic teachers Hannes Leitgeb, Godehard Link and Carlos Moulines for their continous and generous support. This work was supported by the BMBF (grant 01BQ1050, J.N.) and by the DFG (grants NO 1129/1-1 and BO 1734/20-1, J.N.) and the Cluster of Excellence CoTeSys (W.S.).
Further funding (J.N.) was provided by the ERC Advanced Grant IRQUAT, the Spanish MINECO Project No. FIS2013-40627-P and the Generalitat de Catalunya CIRIT Project No. 2014 SGR 966. This paper was presented in part at the Internationl Congress on Mathematical Physics (ICMP) in Santiago de Chile in 2015.
}
\thanks{}
\thanks{\copyright 2016 IEEE. Personal use of this material is permitted. Permission from IEEE must be obtained for all other uses, in any current or future media, including reprinting/republishing this material for advertising or promotional purposes, creating new collective works, for resale or redistribution to servers or lists, or reuse of any copyrighted component of this work in other works.}
\thanks{Digital Object Identifier 10.1109/TIT.2016.2611661}
}

\maketitle

\begin{abstract}
This work is motivated by a question at the heart of unsupervised learning approaches:\\
Assume we are collecting a number K of (subjective) opinions about some event E from K different agents. Can we infer E from them? Prima facie this seems impossible, since the agents may be lying.\\
We model this task by letting the events be distributed according to some distribution p and the task is to estimate p under unknown noise. Again, this is impossible without additional assumptions. We report here the finding of very natural such assumptions - the availability of multiple copies of the true data, each under independent and invertible (in the sense of matrices) noise, is already sufficient:\\
If the true distribution and the observations are modelled on the same finite alphabet, then the number of such copies needed to determine p to the highest possible precision is exactly three! This result can be seen as a counterpart to independent component analysis. Therefore, we call our approach 'dependent component analysis'.\\
In addition, we present generalizations of the model to different alphabet sizes at in- and output. A second result is found: the 'activation' of invertibility through multiple parallel uses.
\end{abstract}
\begin{section}{Introduction\label{sec:introduction}}
Can we know the objective truth about the distribution of a set of events $E$?
\\\\
This question can of course be cast into many different and more precise forms. We will be concerned here with the following version of it: We assume we are collecting a number $K$ of potentially subjective opinions (e.g. from witnesses of some crime) about $E$ from $K$ different agents. A simple sketch of the scenario is given in the following figure.
\definecolor{light-gray}{gray}{0.8}
\definecolor{dark-gray}{gray}{0.4}
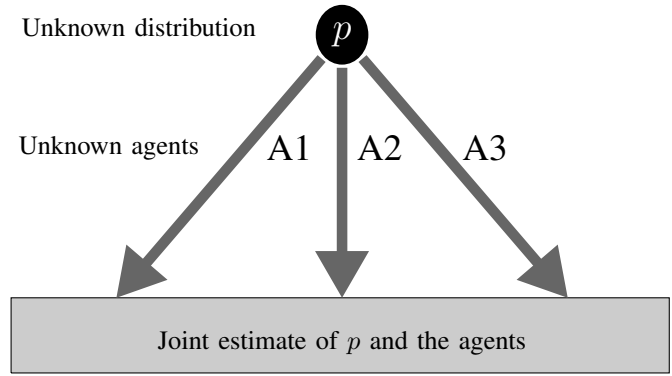
\begin{figure}\begin{tikzpicture}
\node (X) [fill=black,ellipse,inner sep=3.5pt] at (5,6) {\textcolor{white}{{\Large $p$}}};
\node (C1) at (2.3,6.1) {\textcolor{black}{{Unknown distribution}}};
\node (C2) at (1.9,4.5) {\textcolor{black}{{Unknown agents}}};
\node (Cpos)  [inner sep=6pt,anchor=center,align=center] at (5,6.43) {};
\node (A1) [black, ellipse,inner sep=4.5pt] at (4.3,4.5) {{\Large A1}};
\node (A2) [ellipse,inner sep=4.5pt] at (5.5,4.5) {{\Large A2}};
\node (A3) [ellipse,inner sep=4.5pt] at (6.9,4.5) {{\Large A3}};
\draw[fill=light-gray] (0.6,1.5) -- (0.6,2.5) -- (9.35,2.5) -- (9.35,1.5) -- (0.6,1.5);
\node (T)  [fill=light-gray,rectangle,inner sep=6pt] at (5,1.9) {{Joint estimate of $p$ and the agents}};
\path[-triangle 60, color=dark-gray,line width=4pt, shorten <= 1pt] (X) edge (2,2.5);
\path[-triangle 60, color=dark-gray,line width=4pt, shorten <= 1pt] (X) edge (5,2.5);
\path[-triangle 60, color=dark-gray,line width=4pt, shorten <= 1pt] (X) edge (8,2.5);
\end{tikzpicture}\ \\
\caption{Dependent component system with 3 agents}\end{figure}
\ \\
We show in this work that under certain conditions the task is then feasible: We can find out the objective truth from the subjective opinions given that these opinions are only minimally correlated to the true event, and that the different agents give their respective opinions independently from each other (they are not conspiring). It seems reasonable to assume that especially human perception is correlated to a common objective reality, if this objective reality exists. Thus, our model offers a new way of looking at the process in which a multitude of different opinions about the same objective truth can enable an observer having access to all these opinions to actually find the objective truth. We interpret our result as a mathematical statement in the favour of cooperative actions in the following sense: In the task of finding out an objective truth, one could follow the approach of finding an agent that reliably reports only true values, if necessary by building or training it first and then testing it in various situations. Our approach is contrary in nature: We do not seek to find such an agent at all. Instead, we build on the joint use of multiple, uncorrelated observations. In addition, no 'training' is necessary and no assumptions are made on the mutual information between in- and output of the channels. In fact, the mutual information between the true events and the events reported by the agents has to be nonzero but can be arbitrarily small otherwise.
\\\\
From an information-theoretic perspective, which we adopt in this work, the task of finding the true state of some object or process is formulated best in terms of hypothesis testing. A huge amount of fundamental results has been obtained e.g. in \cite{chernoff,blahut,csiszar-longo,kullback}. A simple introduction to basic reasoning in this area can be found in \cite{csiszar-koerner}.\\
However, the task of hypothesis testing is usually formulated such that a direct access to the source, system or process is guaranteed. In the light of developments e.g. in quantum theory or with an eye on extraterrestrial exploration, this seems highly questionable - the system to be observed is usually being observed via its interaction with a measurement apparatus, which in the event gets correlated to the system. The observer ultimately draws his conclusions from the output given to him through the measurement device.\\
One may now argue that the uncertainties in the measurement apparatus could in many situations be circumvented by adjusting it properly. This certainly requires that one makes measurements on already well-known inputs. One readily sees that this argument is circular in nature - how did one come to know these well-known inputs?\\
It seems reasonable to take one step back and consider the task of hypothesis testing under unknown noise. Such noise may for example be introduced through imperfections built in a measurement apparatus, or the lack of knowledge regarding events reported to us via not necessarily trusted agents.
\\\\
The development of a complete theory of dependent components systems is way beyond the scope of this work. Although we aim at formulations in traditional hypothesis testing scenarios, there is one basic question which has to be answered at first and it is this question that we first pose and then solve here:
\begin{center}
\emph{Is a dependent component system invertible?}
\end{center}
Of course, if two different distributions $p$ and $p'$ of the events $E$ could potentially get mapped to one and the same output distribution $q$ by the influence of noise, any approach to differentiate between $p$ and $p'$ would be doomed to fail. Thus, it is of utmost importance to clarify this one point before starting to formulate more elaborate tasks.
\\\\
Before we go into more detail, we now give a first and informal definition of the term 'dependent component analysis'. For simplicity, we will call the systems under consideration dependent component systems ($DCS$). Quite generally, such a system is to be understood as any physical system in a given state $p$, together with a number $K$ of channels (linear, positivity-preserving maps going from the system to their respective output systems).\\
The goal of $DCA$ is to determine, from data taken from all or some of the $K$ channel outputs, the true state $p$ of the system.\\
More specifically we will, throughout this work, assume that the system under consideration is given by a probability distribution $p$ on a finite alphabet $\{1,\ldots,L\}$ which simply labels the events $E$ (without loss of generality the events $E$ are therefore given by natural numbers). Through the time of $n\in\nn$ observations, the system generates the events $(E_1,\ldots,E_n)$ which are distributed independently and identically according to $p$. Each channel receives an exact copy of this sequence and transmits it to the output. The channels are assumed to be memoryless. They act independently from each other.\\
\emph{Our main result in this situation is the following: As long as the set of possible events at the output of each of the channels has the same cardinality as the set of possible events $E$ at the input, the number $K$ of channels satisfies $K\geq3$ and each of them is invertible as a matrix, the distribution $p$ of the events $E$ can be inferred up to a permutation if one knows the distribution of events at the output (which can be approximated to arbitrary precision from observed data due to the assumed structure of the channels and distribution $p$). We additionally prove that even non-invertible channels can be used to obtain this result, if only enough of them are available.}
\\\\
{\bf Outline of the paper.} We first state our main results in Section \ref{sec:main-results}. These clarify when a $DCA$ system can be inverted. We give examples for non-invertible systems as well. Thus an open question remains: Under which circumstances is a $DCA$ system invertible, and can this be detected solely from observations at the output of the system \emph{and} from knowing that it \emph{is} in fact a $DCA$ system?\\
A partial answer to this is given by our Theorem \ref{theorem:inversification}, which states that multiple parallel uses of one and the same channel can be inverted if the inputs are restricted to a certain form, even if the channel itself is non-invertible.\\
The proof of our statements are given in Section \ref{sec:proofs}. In the appendix (Section \ref{sec:appendix}) we provide an additional subsection which highlights the connection to hypothesis testing and clarifies how the overall detection process can be carried out. This connects our approach to \cite{blahut,kullback,csiszar-longo,chernoff} - our work is a first crucial step towards a generalization of hypothesis testing to situations where the test takes place under some additional, unknown noise. Finally in subsection \ref{sec:simpson-paradox} we briefly connect to the Simpson-Yule paradox and the 'conjunctive fork'. Page numbers are as follows:
\setcounter{tocdepth}{1}
\renewcommand*\contentsname{{}}
\begin{center}\begin{minipage}[c]{8cm}{\small\tableofcontents}\end{minipage}\end{center}
\begin{subsection}{Historical notes and connections to other approaches}
The reader interested in the subject will find a multitude of different approaches to systems with additional structure, like the one treated here. For example the approach taken in this work applies to multiple antenna systems and radar as well, and this is an area of research that already reported use of the effect described here as early as 1931 in \cite{beverage-peterson,beverage-peterson-moore}, although the mathematical treatment given in this work differs strongly from these earlier approaches. Our approach is finally able to provide a deeper and very general understanding of the phenomenon from a clean perspective, if only at the price of a finite-dimensional analysis.\\
Another application that seems to fall into the category of dependent component systems is human perception: The different sensing systems (e.g. vision and hearing) can be assumed to be subjected to independent noise most of the time. Anything which affects both vision and hearing at the same time can, according to everyday-experience, in general be identified very precisely.\\
With multiple independent copies, perspectives and opinions on all kinds of subjects via the internet, dependent component analysis (which will usually be abbreviated by $DCA$ in the following) can certainly be applied in data analysis as well, and at least in spirit this effect is exploited for noise estimation in digital images e.g. in the recent work \cite{shih-kwatra-chinen-fang-ioffe}.

As mentioned already, such studies date back at least as far as the 1930's. We therefore confine ourselves here to the mentioning of only a few research areas which we feel are important either from practical or theoretical, if not even philosophical perspectives. We also restrict ourselves to citing only very few published results in these areas, and we picked them such that the references contained therein enable the reader to quickly enter the corresponding field.\\
At first, let us mention the famous $ICA$ (independent component analysis) approach, which can be considered orthogonal in spirit to ours. In $ICA$, the system under consideration consists of $K$ independent parts, and the transformation between the system and the observer is only assumed to be linear.\\
The astonishing result in $ICA$ is, that it is possible to detect both channel and system, up to a permutation, but only from observing the output and from knowing that the system under consideration fulfills above assumptions. For a good introduction to ICA, including its history, see \cite{comon} or \cite{hyaevarinen-oja}.
\\\\
Another branch which has to be mentioned here is the analysis of multichannel systems. An introduction to these topics can be found for example in \cite{viswanathan-varhsney-I} or \cite{blum-kassam-poor-II}. A first paper summarizing different approaches to the topic was published by Brennan as early as 1959 \cite{brennan}. Many contributions from the engineering perspective can be found under the keyword 'diversity combining'.\\
Surprisingly, it seems the situation has never been analyzed in an information-theoretic context. The results which are known to the authors consider several restoration problems, among them image restoration, but do not exploit the specific probabilistic structure itself nor do they consider the various problems arising from different alphabet sizes. Also, it seems to have slipped the attention of earlier research that dependent component systems can, under not too strong assumptions, be inverted. Thanks to the comments of an unknown reviewer, we were made aware of the publication \cite{wosvw-DUDE} that treats the problem of denoising in an information-theoretic framework. In that work, a string $x^n=(x_1,\ldots,x_n)$ representing undistorted information, is subjected to noise acting independently on every symbol $x_i$, producing the output $y^n=(y_1,\ldots,y_n)$. An algorithm (called discrete universal denoiser or DUDE, for short) is presented that, when the parameters of the noise model are known, delivers optimal performance in the recovery of $x^n$ from $y^n$. Of course those parameters may not always be given, in which case the algorithm is completely useless. However, if multiple copies of $x^n$ have been stored in the past and then corrupted by independent noise, our algorithm is able to deliver exactly the noise model that serves as an input to the algorithm presented in \cite{wosvw-DUDE}. The DUDE has been extended to cover situations involving channel uncertainty in \cite{gemelos-sigurjonsson-weissmann}.\\
Our model is further intimately connected to the study of Markov chains (see e.g. the work \cite{ephraim-merhav} for an overview), as our model can be viewed as a hidden ``arbitrarily varying'' Markov model in which the state space of the Markov chain equals its output space, but the information delivered to the observer is not via one and the same noisy channel, but via different channels:
\begin{figure}\begin{tikzpicture}
\node (X) [fill=black,ellipse,inner sep=4.5pt] at (9,6) {\textcolor{white}{{\Large $p$}}};
\node (X1) [fill=black,ellipse,inner sep=4.5pt] at (1,6) {};
\node (X2) [fill=black,ellipse,inner sep=4.5pt] at (4,6) {};
\node (X3) [fill=black,ellipse,inner sep=4.5pt] at (7,6) {};
\path[-triangle 60, color=black,line width=4pt, shorten <= 1pt] (X) edge (X3);
\path[-triangle 60, color=black,line width=4pt, shorten <= 1pt] (X3) edge (X2);
\path[-triangle 60, color=black,line width=4pt, shorten <= 1pt] (X2) edge (X1);
\node (Cpos)  [inner sep=6pt,anchor=center,align=center] at (5,6.43) {};
\node (A1) [black, ellipse,inner sep=4.5pt] at (1.5,4.5) {{\Large $W_3$}};
\node (A2) [ellipse,inner sep=4.5pt] at (4.5,4.5) {{\Large $W_2$}};
\node (A3) [ellipse,inner sep=4.5pt] at (7.5,4.5) {{\Large $W_1$}};
\draw[fill=light-gray] (0.6,1.5) -- (0.6,2.5) -- (9.35,2.5) -- (9.35,1.5) -- (0.6,1.5);
\node (T)  [fill=light-gray,rectangle,inner sep=5pt] at (4.8,1.95) {{Joint estimate of $p$ and the noise models $W_1,W_2,W_3$}};
\path[-triangle 60, color=dark-gray,line width=4pt, shorten <= 1pt] (X1) edge (1,2.5);
\path[-triangle 60, color=dark-gray,line width=4pt, shorten <= 1pt] (X2) edge (4,2.5);
\path[-triangle 60, color=dark-gray,line width=4pt, shorten <= 1pt] (X3) edge (7,2.5);
\end{tikzpicture}\ \\
\caption{Dependent component system seen as Markov chain with different (unknown) noisy channels $W_1,W_2,W_3$ and unknown input distribution $p$.}\end{figure}
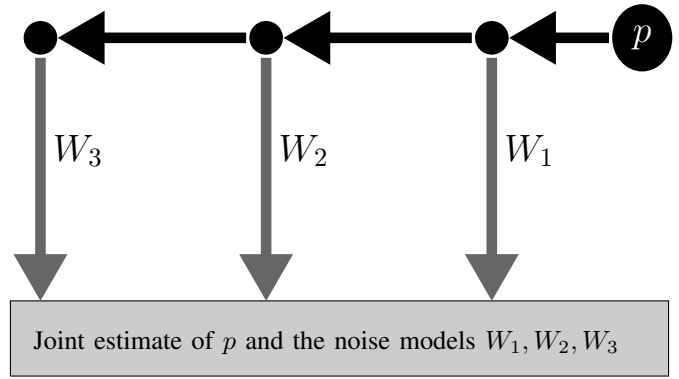
\ \\
After finalization of the key results and statements of this manuscript we became aware that the name ``dependent component analysis'' has actually been used in the literature earlier already (see or example \cite{dca-book} or \cite{li-li-dca} and references therein for an overview), although not in the framework that is treated here. Rather, a multitude of different approaches dealing with estimation of multivariate distributions is presented there. To the author's knowledge, this work is the very first to pose the fundamental question of invertibility together with the question of how many independent observations one has to make in order to invert the system. That the answer to this question is exactly three is a surprising result, which we are at present tempted to not see as a mere artifact.\\
Invertibility of channel matrices has also played a minor role in the work \cite{han-verdu-output} on approximation of output statistics of channels: After developing a theory of resolvability for output statistics of channels, the authors observed that statements about its output statistics translate to statements about the input statistics when the channel is invertible (see the discussion of \cite[Theorem 15]{han-verdu-output}). In a second paper \cite{han-verdu-full-rank-channels}, the authors of \cite{han-verdu-output} explicitly exploited this observation to further develop the theory of resolvability by extending it to input processes of a channel.\\
The type of question we study here is, at least in spirit, similar to the search for informationally complete measurements in quantum information theory \cite{busch,dariano-perinotti-sacchi,dariano-perinotti-sacchi-universal,prugovecki}. Such measurements have the property that they guarantee the possibility to distinguish between the many possible states that a physical system may be in.\\
From the recent work \cite{mueller-masanes} which is inspired among others by results of C.F. von Weizs\"acker \cite{weizsaecker} it is known that the quantum bit space (which can be represented as the unit ball in exactly three real dimensions) and the three-dimensional space structure we are experiencing every day can actually be related by a number of clearly specified reasonable assumptions and logical arguments. We hypothesize that similar arguments should make it possible to connect our findings to the geometry of space.
\end{subsection}
\end{section}
\begin{section}{Notation, motivation and introduction of the framework of multilinear algebra\label{sec:notation-and-conventions}}
This section is split into three subsections. First we define the standard notation, a task which is rather repetitive. Then we give a more precise formulation of our problem statement, using the established notation. We explain which difficulties arise. Then, we introduce the framework of multilinear algebra which we use later to solve the problem.
\begin{subsection}{Notation\label{subsec:notation}}
For a natural number $L$, we define $[L]:=\{1,\ldots,L\}$. The set of permutations on $[L]$ is denoted $S_L$. Given two such sets $[L_1],[L_2]$, their product is $[L_1]\times[L_2]:=\{(l_1,l_2):l_1\in[L_1],\ l_2\in[L_2]\}$. For any natural numbers $n$ and $L$, $[L]^n$ is the $n$-fold product of $[L]$ with itself. The set of real-valued functions on a finite set $[L]$ is $\mathfrak F_L$. The set of probability distributions on a finite set $[L]$ is
\begin{align}
\mathfrak P([L]):=\{p\in\mathfrak F_L:p(i)\geq0\ \forall\ i\in[L],\ \sum_{i=1}^Lp(i)=1\}.
\end{align}
The support of $p\in\mathfrak P([L])$ is $\supp(p):=\{i\in[L]:p(i)>0\}$. A very important subset of elements of $\mathfrak P([L])$ is the set of its extreme points, the Dirac-measures: for $i\in[L]$, $\delta_i\in\mathfrak P([L])$ is defined through $\delta_i(j)=\delta(i,j)$, where $\delta(\cdot,\cdot)$ is the usual Kronecker-delta symbol. Closely related to the latter is the set of indicator functions on $[L]$. These are defined for subsets $S\subset[L]$ via $\eins_S(i)=1$ if and only if $i\in S$ and $\eins_S(i)=0$ else.\\
Two subsets of $\mathfrak P([L])$ which are of great importance in our case are
\begin{align}
\mathfrak P_>([L])&:=\{p\in\mathfrak P([L]):p(i)>0\ \forall\ i\in[L]\}\\
\mathfrak P^\downarrow([L])&:=\{p\in\mathfrak P([L]):p(1)>\ldots>p(L)\}.
\end{align}
The main task that we will be dealing with is that of \emph{distinguishing} between elements $p,p'\in\mathfrak P([L])$. While there are many ways of doing this, we would like to single out two of them here: First, we may use (in principle for any number $k\in\mathbb R$, though we will make use only of $k=1$ and $k=2$ here) the $k$-norms $\|p-p'\|_k:=(\sum_{i=1}^L|p(i)-p'(i)|^k)^{1/k}$. If $k=1$, we will omit the subscript in the following.\\
Second, one may use the relative entropy or Kullback-Leibler distance: $D(p\|p')$ is defined as $D(p\|p'):=\sum_{i=1}^Lp(i)\log(p(i)/p'(i))$ if $p'(i)=0\Rightarrow p(i)=0\ \forall i\in[L]$ and $D(p\|p'):=+\infty$ if this is not the case.\\
The noise that complicates the task of estimating $p$ is modelled by matrices $W$ of conditional probability distributions $(w(i|j))_{i,j=1}^{L',L}$ whose entries are numbers in the interval $[0,1]$ satisfying, for all $j\in[L]$, $\sum_{i=1}^Lw(i|j)=1$. The set of all such matrices is denoted $\mathcal W([L],[L])$. Such matrices are, using standard terminology of Shannon information theory, equivalently called a ``channel''.\\
We will later relate these probabilistic concepts to linear algebra in a more explicit fashion. In order to motivate this approach, we first restate the problem using the rather generic language that we introduced so far:
\end{subsection}
\begin{subsection}{Problem statement (standard formalism)\label{subsec:problem-statement(standard-formalism)}}
Let $L,K\in\nn$ be given. To a given choice $W_1,\ldots,W_K\in\mathcal W([L],[L])$ of channels and $p\in\mathfrak P_>([L])$ we let $q\in\mathfrak P([L]^K)$ be the distribution defined by setting, for all $y_1,\ldots,y_K\in[L]$,
\begin{align}
q(y_1,\ldots,y_K)=\sum_{x\in[L]}p(x)\prod_{k=1}^Kw_k(y_k|x).
\end{align}
The question we ask is, for what other choices $p'\in\mathfrak P([L])$ and $W_1,\ldots,W_k\in\mathcal W([L],[L])$ the equalities
\begin{align}\label{eqn:primed-quantities}
q(y_1,\ldots,y_K)=\sum_{x\in[L]}p'(x)\prod_{k=1}^Kw'_k(y_k|x)
\end{align}
can hold for all $y_1,\ldots,y_k\in[L]$. It is clear that the answer is completely unsatisfying for $K=1$: For example the choice of the distribution $p'$ satisfying \eqref{eqn:primed-quantities} is completely independent from that of $p$, since it is always possible to choose an appropriate $W'_1$ such that \eqref{eqn:primed-quantities} is fulfilled. Later in example \ref{example:K=2} we explain shortly why the answer for $K=2$ turns out to be not very satisfying as well. With the help of our Theorem \ref{theorem:uniqueness-of-solution} however we can demonstrate that, starting from $K=3$, the solutions to the $L^K$ equations (one for each choice of $y_1,\ldots,y_K\in[L]$):
\begin{align}\label{eqn:basic-question}
\sum_{x\in[L]}p(x)\prod_{k=1}^Kw_k(y_k|x)=\sum_{x\in[L]}p'(x)\prod_{k=1}^Kw'_k(y_k|x)
\end{align}
can be written as follows: Take any choice $p,W_1,\ldots,W_K$. For every permutation $\tau:[L]\to[L]$, the distribution $p'$ defined by $p'(i):=p(\tau(i))$ and channels $W_k'$ defined by $w_k'(j|i):=w_k(j|\tau^{-1}(i))$ are a solution to the equation \eqref{eqn:basic-question}, and for a fixed choice of $p$ and $W_1,\ldots,W_K$ these are the only solutions.\\
Proving this result turned out to be a challenging task: Starting with restricted models where $L=2$ and all channels $W_k$ are restricted to lie in the class of binary symmetric channels $W_k=BSC_{s_k}$ with unknown probability $s_k$ of causing a bit to be flipped, we realized that the polynomial equation in the variables $p(1),s_1,s_2$, and $p'(1),s_1',s_2'$ arising from \eqref{eqn:basic-question} has only finitely many solutions. We were also able to picture the manifolds arising from a fixed choice of $p(0)$ and variations of $s_1,s_2$ and realized that they were forming a two-dimensional manifold.
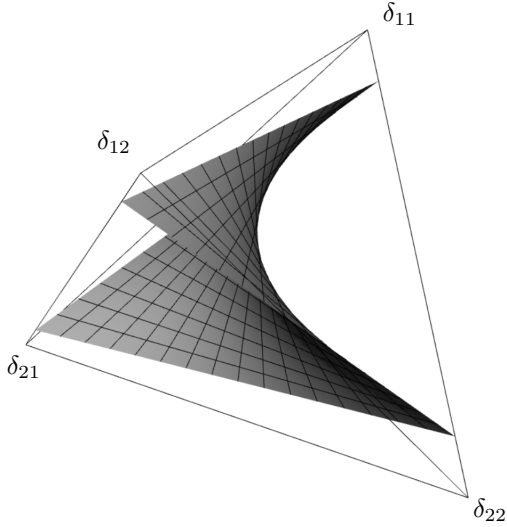
\begin{figure}
\hspace*{-15em}
\begin{tikzpicture}
\includegraphics[width=6cm]{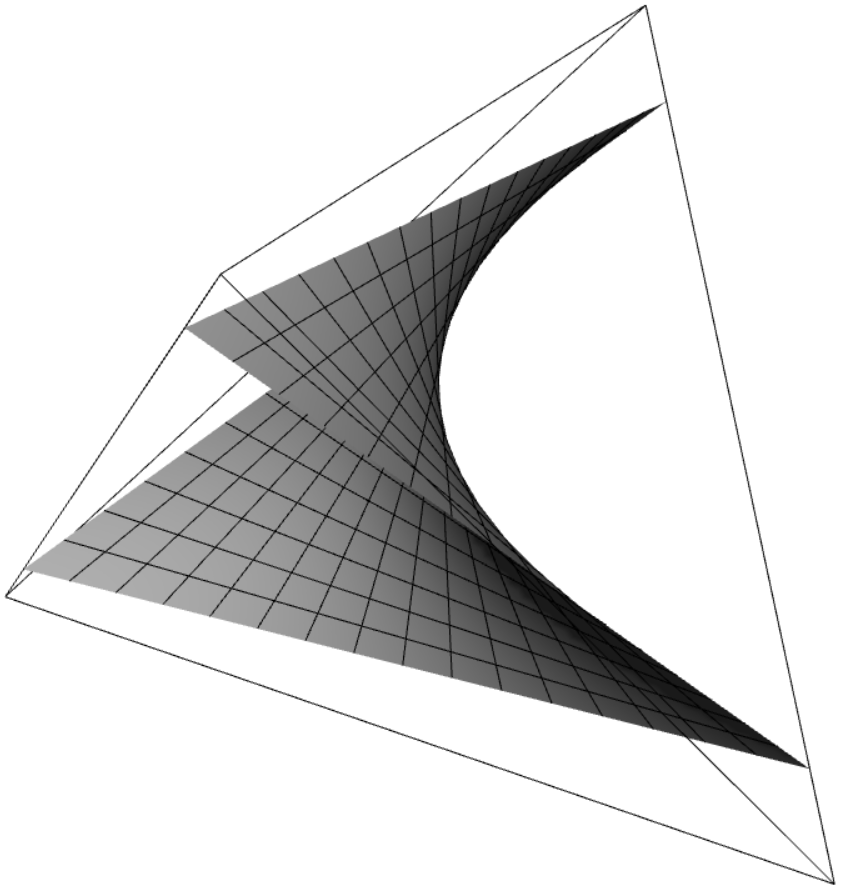}
\node (X) at (0.2,-0.1) {$\delta_{22}$};
\node (X) at (-6,1.8) {$\delta_{21}$};
\node (X) at (-4.8,4.8) {$\delta_{12}$};
\node (X) at (-1,6.5) {$\delta_{11}$};
\end{tikzpicture}
\caption{\label{fig:bsc}Probability simplex over the alphabet $\{11,12,211,22\}$. The $\delta_{ij}$ denote the Dirac-distributions corresponding to the respective letters in the alphabet. The two-dimensional manifold arises from setting $p(1)=0.12$ and $K=2$ in \eqref{eqn:basic-question} and letting the channels $w_1$ and $w_2$ be binary-symmetric channels with flip probabilities $(s_1,s_2)\in[0,1]\times[0,1]$.}
\end{figure}
This motivated further study of the question. Unfortunately we were not able to solve the resulting polynomial equations in the many variables that arise as soon as the model becomes even slightly more complex. From the above picture it became clear that $K=2$ would not be sufficient for more complex noise models. In fact, counting parameters led us to conclude the following: As any channel $W$ introduces $(L-1)\cdot L$ variables, and $p$ another $L-1$ variables, we could only hope to gain insightful solutions whenever the total number $K\cdot(L-1)\cdot L+L-1$ of variables was smaller than or equal to the dimension of the convex set $\mathfrak P([L])^K$, which equals $L^K-1$. For natural numbers $L,K$, the inequality
\begin{align}\label{eqn:parameter-counting}
L^K\geq K\cdot(L-1)\cdot L+L
\end{align}
however is easily seen to hold true for all $K\geq3$. Unfortunately, already the binary case $L=2$ with arbitrary channels $W_1,W_2,W_3,W_1',W_2',W_3'\in\mathcal W(2,2)$ introduces an untractable total number of $2\cdot6+2=14$ variables, and no results concerning polynomials of the specific form introduced by our model could be found. After numerous attempts we finally reformulated the problem using the formalism that was developed during the evolution of quantum information theory over the last 20 years, and this approach finally turned out to be fruitful and helped us to reduce the problem to a small number of rather simple observations. The necessary notation for the reformulation of the problem is given in the following subsection.
\end{subsection}
\begin{subsection}{Multilinear framework}
We will from now on consider any $\mathfrak P([L])$ as being embedded into $\mathbb R^L$ through the bijective map $\Phi:\mathfrak P([L])\to\mathbb R^L$ defined through $\Phi(\delta_i):=e_i$, where $\{e_i\}_{i=1}^L$ is any fixed orthonormal basis of $\mathbb R^L$. The image $\Phi(\mathfrak P([L]))$ naturally generates its $L-1$-dimensional supporting hyperplane within $\mathbb R^L$.\\
This embedding allows a natural use of matrix calculus. Since we will be dealing with composite systems throughout, an important part of our work requires basic results from multilinear algebra. We will now introduce these for bipartite systems, the generalization to the multipartite case is straightforward.\\
Throughout, we use one fixed basis $\{e_i\}_{i=1}^L$ for $\mathbb R^L$. $L$ times $L'$ matrices are thought of as linear maps from $\mathbb R^L$ to $\mathbb R^{L'}$ via their action in this basis. The set of linear maps from $\mathbb R^L$ to $\mathbb R^{L'}$ (matrices with $L'$ rows and $L$ columns) is denoted $M(L,L')$, the group of invertible matrices (if $L=L'$) is written $Gl(L)$. The range of a matrix $M\in M(L,L')$ is $\ran M:=\{Mx:x\in\mathbb R^L\}$, and its kernel $\ker M$ (as usual) the set $\{x\in\mathbb R^L:Mx=0\}$.\\
The scalar product $\langle\cdot,\cdot\rangle$ on $\mathbb R^L\times\mathbb R^L$ is the standard one: $\langle e_i,e_j\rangle=\delta(i,j)$.
If needed, we will represent an $L'\times L$ matrix by using the matrix basis $\{E_{i,j}\}_{i,j=1}^{L',L}$ which is defined by $E_{i,j}e_k=\delta(k,j)e_i$ $\forall i\in L',\ k,j\in L$. The identity matrix on $\mathbb R^L$ (which is at the same time a channel from $[L]$ to $[L]$) is written as $\eins=\sum_{i=1}^L E_{ii}$. Care will be taken that it does not get confused with indicator functions (symbols of the form $\eins_S$) that we defined earlier.\\
Without going into any further detail, we introduce the tensor product of $\mathbb R^L$ with $\mathbb R^K$ in a very straightforward manner by setting
\begin{align}
\mathbb R^L\otimes\mathbb R^K:=\linspan\{e_i\otimes e_k\}_{i=1,k=1}^{L,K}.
\end{align}
This allows us to define general 'product vectors' of two vectors $u=\sum_{i=1}^Lu_ie_i$ and $v=\sum_{i=1}^Kv_ie_i$ by $u\otimes v:=\sum_{i,j=1}^{L,K}u_iv_je_i\otimes e_j$. The vector space $\mathbb R^L\otimes\mathbb R^K$ inherits the scalar products from $\mathbb R^L$ and $\mathbb R^{L'}$ by the formula $\langle u\otimes v,x\otimes y\rangle:=\langle u,x\rangle\langle v,y\rangle$. The corresponding matrix spaces are denoted $M((L,K),(L',K'))$. Given $A\in M(L,L')$ and $B\in M(K,K')$, we define $A\otimes B\in M((K,L),(K',L'))$ through its action on product vectors:
\begin{align}
(A\otimes B)(u\otimes v):=(Au)\otimes(Bv).
\end{align}
This leads to the following set of rules:
\begin{align}
&\forall A,B\in M(L),\ C,D\in M(L):\nonumber\\
&\qquad\qquad(A+B)\otimes(C+D)=\nonumber\\
&\qquad\qquad\qquad A\otimes C+A\otimes D+B\otimes C+B\otimes D,\\
&\forall A\in Gl(K),\ B\in Gl(L):\nonumber\\
&\qquad\qquad(A\otimes B)^{-1}=A^{-1}\otimes B^{-1}\\
&\forall A\in M(K),\ B\in M(L),\ \alpha\in\mathbb C:\nonumber\\
&\qquad\qquad\alpha(A\otimes B)=(\alpha A)\otimes B=A\otimes(\alpha B)\\
&\qquad\qquad\eins_{\mathbb R^K\otimes\mathbb R^L}=\eins_{\mathbb R^K}\otimes\eins_{\mathbb R^L}.
\end{align}
In order to simplify notation later we will use, for $u\in\mathbb R^L$ and $n\in\nn$, the shorthand $u^{\otimes n}:=u\otimes\ldots\otimes u$ for the $n$-fold tensor product of $u$ with itself. Accordingly, for $A\in M(L,K)$ we write $A^{\otimes n}$ to denote the $n$-fold product $A\otimes\ldots\otimes A$. A very important object we shall encounter is the vector $v_+:=\sum_{i=1}^Le_i\otimes e_i\in\mathbb R^L\otimes\mathbb R^L$. It has the important property that
\begin{align}
(A\otimes\eins)v_+=(\eins\otimes A^\top)v_+\qquad\forall A\in M(L),
\end{align}
where the matrix transposition $A\mapsto A^\top$ is defined in the basis $\{e_i\}_{i=1}^L$.\\
An important operation on a composite system described by the set of probability distributions $\mathfrak P([K]\times[L])$ is the partial trace $\tr_{1}:\mathbb R^K\otimes\mathbb R^L\to\mathbb R^L$ which 'forgets' the information stored in the first system modelled over $\mathbb R^K$ by summing over it: For $a=\sum_{i,j=1}^{K,L}a_{i,j}e_i\otimes e_j$, it is defined by
\begin{align}
\tr_{1}(a):=\sum_{i,j=1}^{K,L}a_{i,j}e_j.
\end{align}
This operation has the following nice property: If $v=\sum_{i,j=1}^{K,L}p(i,j)e_i\otimes e_j$ for some $p\in\mathfrak P([K]\times[L])$, then
\begin{align}
\tr_{1}(v)=\sum_{i=1}^L\left(\sum_{j=1}^Kp(i,j)\right)e_i,
\end{align}
which is clearly just the image of one of the marginal distributions of $p$ under the (injective) map $\Phi$. The notion easily extends to multi-partite systems $\mathbb R^{L_1}\otimes\ldots\otimes\mathbb R^{L_m}$ and defines $\tr_i$ for every $i\in[m]$ and, more generally, $\tr_A$ for every $A\in[m]$. Note that if only one system being modelled by $p\in\mathfrak P([L])$ is present we can as well abbreviate $\tr_1$ by $\tr$ and then it holds $\tr_(p)=\sum_{i=1}^Lp(i)=1$.\\
This way we have come back to probability distributions and are now in the position to embed the most natural linear maps on them into our framework: channels. A channel is a positivity and trace preserving linear map $W:\mathfrak P([L])\to\mathfrak P([L'])$, where $L,L'\in\nn$ are arbitrary. We may think of $W$ as a matrix defined by its entries $w_{ij}:=W(\delta_j)(i)$. If necessary and unambiguous, we shall also write $w(i|j):=w_{ij}$.\\
An important example of a channel is a permutation matrix $\tau\in S_L$: Its natural action as $\tau\in\mathcal W([L],[L])$ is via $\tau(p)(i):=p(\tau^{-1}(i))$.\\
Clearly, every channel is completely represented by the matrix with entries $w_{ij}$ and application of $W$ to a probability distribution $p\in\mathfrak P([L])$ is equivalent to applying the matrix defined via its matrix entries $w_{ij}:=w(i|j)$ to the vector $\sum_{i=1}^Lp(i)e_i$.\\
During our proofs, we will not necessarily always be working with channels, but rather with matrices. We will therefore spend a few more words on this connection.\\
It is clear that $\mathfrak P([L])\subset\mathbb R^L_1:=\{\sum_{i=1}^Lv_ie_i\in\mathbb R^L:\sum_{i=1}^Lv_i=1\}$. Therefore, we have the implication $W\in\mathcal W([L],[L])\ \Rightarrow\ W(\mathbb R^L_1)\subset \mathbb R^L_1$, and in case that $W$ is invertible this lets us conclude that $W^{-1}(\mathbb R^L_1)=\mathbb R^L_1$.\\
We see that, as long as we restrict our analysis to matrices which are composed of channels or inverses of channels, all we need to take care of is their action on $\mathbb R^L_1$.\\
In addition, a channel $W\in\mathcal W([L],[L])$ is invertible if and only if the corresponding matrix $W\in M(L,L)$ is invertible. While it is clear that a non-invertible channel has a non-invertible matrix associated to it, the other direction of the claim can be established as follows:\\
Let for some real numbers $\gamma_1,\ldots,\gamma_L$, not all of which are zero, $\sum_{i=1}^L\gamma_iW(\delta_i)=0$. Set $g:=\sum_{i=1}^L\gamma_i$. If $g\neq0$ we may set $\bar\gamma_i:=\gamma_i/g$ for each $i\in[L]$ and conclude that $0=\sum_{i=1}^L\bar\gamma_iW(\delta_i)$, in contradiction to $W(\mathbb R_1^L)\subset\mathbb R_1^L$. If $g=0$ we define the number $\gamma_{\max}:=\max_{i\in[L]}|\gamma_i|$. For every $\alpha\in\mathbb R$, we introduce the vector $\gamma_\alpha:=\sum_{i=1}^L(\alpha\cdot\gamma_i+L^{-1})\delta_i$. For each $\alpha$ it is clear that $\gamma_\alpha\in \mathbb R_1^L$. We additionally have
\begin{align}
(\gamma_\alpha)_i&=\alpha\cdot\gamma_i+L^{-1}\\
&\geq L^{-1}-|\alpha|\cdot\gamma_{\max}\\
&\geq0
\end{align}
whenever $\alpha\in[0,\tfrac{1}{L\cdot\gamma_{\max}}]$, meaning that $\gamma_\alpha\in\mathfrak P([L])$ for that range. By assumption $W(\alpha\cdot\gamma)=0$ for all $\alpha$ and this implies that especially $W(\gamma_\alpha)=W(\gamma_0)$ for all $\alpha\in[0,\tfrac{1}{L\cdot\gamma_{\max}}]$. But at least one of the $\gamma_i$ is nonzero, thus $\gamma_\alpha\neq \gamma_{\alpha'}$ for all $\alpha,\alpha'\in[0,\tfrac{1}{L\cdot\gamma_{\max}}]$ for which $\alpha\neq\alpha'$. This is in contradiction to the assumed invertibility of $W$ as a map from $\mathfrak P([L])$ to $\mathfrak P([L])$.
\end{subsection}
\end{section}
\begin{section}{Definitions\label{sec:definitions}}
We are now able to formalize our problem using the language of multilinear algebra. Throughout our discussion, we will assume the existence of a probability distribution $p$ on some finite set $[L]$ that models the 'true' state of a physical system which emits signals $i\in[L]$ independently and identically distributed according to $p$.\\
Since $p$ is fixed, it makes little sense to talk about events which do not happen according to $p$. Therefore, we will always assume that $p(i)>0$ holds for all $i\in[L]$. In other words: $p\in\mathfrak P_>([L])$.\\
On the other hand, this requires us to 'learn' the parameter $L$ of the system as well. We therefore list three different scenarios: We start with the case $L=L'$, which is central to the whole discussion and delivers the necessary tools for a discussion of the other cases, namely $L'<L$ and $L'>L$.\\
We will now list the definitions that we need in order to state our results. First, a technical thing:
\begin{definition}
Given $p\in\mathfrak P([L])$ we denote by $p^{(K)}$ the distribution
\begin{align}
p^{(K)}:=\sum_{i=1}^Lp(i)\delta_i^{\otimes K}.
\end{align}
For any fixed $[L]$, the set of all such $p^{(K)}$ is
\begin{align}
\mathfrak P^{(K)}([L]):=\{p^{(K)}:p\in\mathfrak P([L])\}.
\end{align}
\end{definition}
The main definition is the following.
\begin{definition}[Dependent component system (DCS)]
For given natural numbers $L,L'$ and $K$, we define For given natural numbers $L,L'$ and $K$, we define $DCS(L,K,L')$ to be the set of all $(p,W_1,\ldots,W_K)$ such that $p\in\mathfrak P([L])$ and $W_1,\ldots,W_K\in\mathcal W([L],[L'])$.
This is the set of all dependent component systems. Any element of it is represented (uniquely only if $p\in\mathfrak P_>([L])$) by the distribution
\begin{align}
\left(\eins\otimes\bigotimes_{i=1}^KW_i\right)p^{(K+1)}\in\mathfrak P([L]\times[L']^K).
\end{align}
\end{definition}
In order to shorten notation we will usually write sentences like 'let a $DCS(L,K,L')$ be given', implying that the mathematical object under study is a system $\mathcal S\in DCS(L,K,L')$.\\
For a more thorough analysis of dependent component systems we need additional definitions:
\begin{definition}
For $L,L',K\in\nn$ we define the $DCS(L,K,L')$ surface to be
\begin{align}
\{(\otimes_{i=1}^KW_i)p^{(K)}:(p,\mathbf W)\in DCS(L,K,L')\},
\end{align}
where $(p,\mathbf W)=(p,W_1,\ldots,W_K)$. As an important subset of this surface we consider the $FR(L,K,L')$ of all distribtions $(\otimes_{i=1}^K W_i)p^{(K)}\in\mathfrak P([L']^K)$ such that $p\in\mathfrak P_>([L])$ and $\ran(W_1)=\ldots=\ran(W_K)=\mathbb R^{L'}$. This is the set of those points which are generated by full-range channels and strictly positive distributions.\\
In case $L'\leq L$ we are going to need the subset $FRSK(L,K,L')$ of all distributions $(\otimes_{i=1}^K W_i)p^{(K)}\in\mathfrak P([L']^K)$ such that $p\in\mathfrak P_>([L])$ and both $\ran(W_i)=\mathbb R^{L'}$ and $\ker(W_i)=\ker(W_j)$ for all $i,j\in[K]$.\\
This subset of $FR(L,K,L')$ consists of all those points on the $DCS(L,K,L')$ surface which are generated by strictly positive probability distributions and $K$ full range channels such that all of them have the same kernel.
\end{definition}
\begin{remark}
Of course, $FR(L,K,L')=\emptyset$ for $L'>L$.
\end{remark}
Let us have a short look at an insightful example.
\begin{example}[$FR(L,2,L)$\label{example:K=2}]
Let $q\in\mathfrak P([L]^2)$. It can obviously be written as $q(i,j)=p(i)r(j|i)$. If there is no such decomposition such that $\supp(p)=[L]$ and $r$ is invertible, then we may choose an $\eps>0$ as small as we like and an invertible $r'$ together with $\supp(p')=[L]$ such that $q'$ defined via $q'(i,j):=p'(i)r'(j|i)$ satisfies $\|q'-q\|\leq\eps$. Take $(p',Id,r')$ as a $DCS(L,2,L)$ system. Then
\begin{align}
(r'\otimes Id){p'}^{(2)}&=\sum_{i=1}^lp'(i)\sum_{j,k=1}^Lr'(j|i)\delta(k,i)\delta_j\otimes\delta_k\\
&=\sum_{i,j=1}^Lr'(j|i)p'(i)\delta_j\otimes\delta_i\\
&=q'.
\end{align}
It follows that $FR(L,2,L)$ is dense in $\mathfrak P([L]^2)$. The same holds true for $K=1$.
\end{example}
This example clearly demonstrates that practically every distribution in $\mathfrak P([L]^2)$ can arise as an output of a dependent component system. This serves as another motivation to look for solutions to our problem statement \eqref{eqn:basic-question} only for $K\geq3$, thus affirming the intuition that one gets from parameter counting as in \eqref{eqn:parameter-counting}.
\end{section}
\begin{section}{Main results, examples\label{sec:main-results}}
Our main result is the following Theorem \ref{theorem:uniqueness-of-solution}, which has to be read with the following in mind: If $W_1,\ldots,W_K,V_1,\ldots,V_K\in\mathcal W([L],[L])$ are all invertible as matrices and $X_i:=V_i^{-1}\circ W_i$ ($i=1,\ldots,K$), then the matrices $X_i\in M(L,L)$ are invertible and they map $\mathbb R^L_1$ to $\mathbb R^L_1$. Let the set of all matrices which map $\mathbb R_1^L$ to itself be $Gl_1(L)$. Another way of writing this is to set $Gl_1(L):=\{X:\sum_{j=1}^LX_{ji}=1\ \forall\ i\in[L]\}$. The exact connection between Theorem \ref{theorem:uniqueness-of-solution} and our initial question \eqref{eqn:basic-question} is explained in detail at the end of the proof of Theorem \ref{theorem:uniqueness-of-solution}.
\begin{theorem}\label{theorem:uniqueness-of-solution}[Uniqueness of Solution for $L=L'$]
Let $K\in\nn$ satisfy $K\geq3$. Let $p\in\mathfrak P_>([L])$. There are exactly $L!$ tuples $(X_1,\ldots,X_L,p')$ of matrices $X_1,\ldots,X_K\in Gl_1(L)$ and probability distributions $p'\in\mathfrak P_>([L])$ satisfying the equation
\begin{align}\label{eqn:invertibility-L-L}
\sum_{i=1}^Lp(i)\delta_i^{\otimes K}=\left(X_1\otimes\ldots\otimes X_K\right)\left(\sum_{i=1}^Lp'(i)\delta_i^{\otimes K}\right).
\end{align}
These are as follows: For every $\tau\in S_L$, the matrices $X_1=\ldots=X_K=\tau^{-1}$ and $p'=\tau(p)$ solve (\ref{eqn:invertibility-L-L}), and these are the only solutions.\\
As a consequence, the function $\Theta:\mathfrak P([L])\times\mathcal W([L],[L])^{K}\to\mathfrak P([L]^K)$ defined by
\begin{align}\label{eqn:def-of-theta}
\Theta[\left(p,T_1,\ldots,T_K\right)]:=\left(\bigotimes_{i=1}^KT_i\right)\left(\sum_{i=1}^Lp(i)\delta_i^{\otimes K}\right)
\end{align}
is invertible if restricted to the right subset: there exists a function $\Theta':DCS(L,K,L)\to\mathfrak P^\downarrow([L])$ that has the property
\begin{align}\label{eqn:inversion-property-of-Theta'}
\Theta'(\Theta[(p,T_1,\ldots,T_K)])=p,
\end{align}
for all $p\in\mathfrak P^\downarrow([L])\cap\mathfrak P_>([L])$ and those $(T_1,\ldots,T_K)$ for which every $T_i$, $i\in[K]$, is invertible. In addition to that, there exists a second function $\Theta'':DCS(L,K,L)\to\mathfrak P^\downarrow([L])\times\mathcal W([L],[L])^K$ such that
\begin{align}\label{eqn:inversion-property-of-Theta''}
\Theta''(\Theta[(p,T_1,\ldots,T_K)])=(p,T_1,\ldots,T_K)
\end{align}
for all $p\in\mathfrak P^\downarrow([L])\cap\mathfrak P_>([L])$ and those $(T_1,\ldots,T_K)$ for which every $T_i$, $i\in[K]$, is invertible.
\end{theorem}
\begin{remark} The following Theorem \ref{theorem:the-perfect-conspiracy} shows that $\Theta'$ has the inversion property (\ref{eqn:inversion-property-of-Theta'}) even for all $p\in\mathfrak P^\downarrow([L])$ and those $T_1,\ldots,T_K$ that are invertible if restricted to $\linspan(\{e_i:p(i)>0\})$.\\
We will give a proof of this theorem for the interesting case $K=3$ only, the general case offers no increase in insight.\\
It will become apparent from the proof that the theorem can be extended to the case where $p,p'$ are not necessarily probability distributions (in which case the $X_i$ will not necessarily be permutations any more).
\end{remark}
In addition, we provide a generalization to the case where the input alphabet is strictly larger than the output alphabet. This situation should be considered the generic case in all sensor networks that involve a digital-analog converter.
\begin{theorem}[The perfect conspiracy]\label{theorem:the-perfect-conspiracy}
Let $L>L'$ and $K\geq3$. Then
\begin{align}
FR(L',K,L')\cap FR(L,K,L')=FRSK(L,K,L').
\end{align}
\end{theorem}
\begin{remark}\label{remark:onL>L'}
The implication of the theorem is the following important 'rule of thumb': If you observe three outputs of dimension $L'$, and you can verify that your observed distribution $q$ is in $FR(L',K,L')$ then either $L=L'$ or there is a 'perfect conspiracy' between all the channels in the sense that they all delete the same information.\\
If you are happy that there are no apparent contradictions in your system, you may just leave it the way it is and conclude that the truth is given by some $p^\backprime$, which may e.g. be given by a projection of the true $p$ onto some hyperplane within $\mathfrak P([L])$.\\
If on the contrary you are a wary individual, you may always add new channels to the system, suspecting that in fact $L'>L$ and that it will be possible to find a channel which does not participate in the perfect conspiracy.
\end{remark}
We now consider the case $L'>L$, which can also be interpreted as generalization of Theorem \ref{theorem:uniqueness-of-solution} to the singular cases (e.g. those cases where $p\notin\mathfrak P_>([L])$).
\begin{theorem}[The case $L'>L$]\label{theorem:the-case-L'<L}
Any $DCA(L,K,L')$ system with at least three channels and $L'>L$ is invertible up to a permutation on $[L]$.
\end{theorem}
All the previous results are exact, and yet leave us unsatisfied: Assume we observe a set of $K$ outputs of channels with output alphabet $[L']$, restrict our attention e.g. to all the triples of $3$ subsystems and from Theorem \ref{theorem:the-perfect-conspiracy} we infer that $L'<L$ has to hold (because the observed outputs may sometimes not be in $FR(L',K,L')$).\\
Then how can we use this knowledge to calculate $p$?\\
More specifically, is there any hope that a system of $K$ channels $W_1,\ldots,W_K\in\mathcal W([L],[L'])$ can be invertible although $L'<L$ holds? Although we are not yet able to give a complete solution to this question including dependencies of the parameters $K,L,L'$ and potentially necessary assumptions concerning $p$ and the channels $W_1,\ldots,W_K$, we can already answer it in the affirmative.\\
To this end, consider a specific example: A channel $W\in\mathcal W([3],[2])$. The generic situation we encounter in this case will be that
\begin{align}
W(\delta_3)=\lambda W(\delta_1)+(1-\lambda)W(\delta_2),\ \ W(\delta_1)\neq W(\delta_2)
\end{align}
for some $\lambda\in[0,1]$ and up to a permuation on $[3]$. We would like to find out now whether the three vectors $\{W(\delta_i)\otimes W(\delta_i)\}_{i=1}^3$ form a linearly independent set. If that is so, their supporting hyperplane has dimension two, just like $\mathfrak P([3])$. Hence, $W\otimes W$ would be invertible as a map from $\mathfrak P^{(2)}([3])$ to its image in $\mathfrak P([2]^2)$.\\
Assume that, on the contrary, there are $\gamma_1,\gamma_2,\gamma_3\in\mathbb R$ not all of which are zero and such that
\begin{align}
0=\sum_{i=1}^3\gamma_iW(\delta_i)\otimes W(\delta_i)
\end{align}
holds. This is (introducing the abbreviations $w_1:=W(\delta_1),\ w_2:=W(\delta_2)$) equivalent to
\begin{align}
0&=(\gamma_1+\lambda^2\gamma_3)w_1^{\otimes 2}+(\gamma_2+(1-\lambda)^2\gamma_3)w_2^{\otimes 2}\nonumber\\
&\qquad\qquad+\gamma_3\lambda(1-\lambda)(w_1\otimes w_2+w_2\otimes w_1).
\end{align}
In case that $\lambda\notin\{0,1\}$ this implies $\gamma_3=0$, which then leads to $\gamma_1=\gamma_2=0$ as well and the desired result is proven by contradiction.\\
If, however, $\lambda\in\{0,1\}$ holds then the above argument does not work. Then, it is even true that $W\otimes W$ is not invertible, since (w.l.o.g. $\lambda=0$)
\begin{align}
W\otimes W(p^{(2)})&=p(1)W(\delta_1)\otimes W(\delta_1)\nonumber\\
&\qquad+(p(2)+p(3))W(\delta_2)\otimes W(\delta_2).
\end{align}
Only in that case is every information about the difference between $2$ and $3$ destroyed at the output and impossible to recover!\\
Let us now become slightly more general. Consider $W\in\mathcal W([L],[L'])$ and assume that the $W(\delta_i)$, $i=1,\ldots,L'$, are pairwise different. We ask, when exactly can $W^{\otimes K}$, restricted to $\mathfrak P^{(K)}([L])$ be invertible? A partial answer is the following theorem:
\begin{theorem}\label{theorem:inversification}
Let $W\in\mathcal W([L],[L'])$ satisfy $W(\delta_i)\neq W(\delta_j)$ for all $i\neq j\in[L]$. Then $K\geq L(L'-1)$ is sufficient for $W^{\otimes K}$ to be invertible as a map from $\mathfrak P^{(K)}([L])$ to $\mathfrak P([L']^{\otimes K})$.
\end{theorem}
We would like to express our belief that this is at the same time already optimal through the following conjecture:
\begin{conjecture}
Under the preliminaries of Theorem \ref{theorem:inversification} we have the following: If $K<L(L'-1)$, then there is always a channel $W\in\mathcal W([L],[L'])$ satisfying $W(\delta_i)\neq W(\delta_j)$ for all $i\neq j\in[L]$ and such that $W^{\otimes K}$ is not invertible as a map from $\mathfrak P^{(K)}([L])$ to $\mathfrak P([L']^{\otimes K})$.
\end{conjecture}
How do we solve the problem of inverting a $DCA$ system concretely? Given that we estimated $\hat q\in\mathfrak P([L']^K)$, an easy solution which can be computed efficiently is the convex optimization problem
\begin{align}
\arg\min_{W_1,\ldots,W_K,p}D((\otimes_{i=1}^KW_i)p^{(K)}\|\hat q).
\end{align}
If $\hat q$ is indeed an output distribution of a $DCA$ system, above algorithm will return the system (up to a permutation). The ambiguities in the solution can be reduced by optimizing not over all $p\in\mathfrak P([L])$, but only over $\mathfrak P^\downarrow([L])$.\\
Of course, it is not at all necessary to use Kullback-Leibler divergence in the above, one could as well use any norm $\|\cdot\|$ and instead compute \begin{align}
\arg\min_{W_1,\ldots,W_K,p}\|(\otimes_{i=1}^KW_i)p^{(K)}-\hat q\|.
\end{align}
For practical purposes, it may even be useful to use a smooth quantity like $\|\cdot\|_2^2$ defined by $\|x\|_2:=\sqrt{\sum_{i=1}^{d}x_i^2}$, where $d=K\cdot L$ in our application and we employ the embedding $\phi$ of $\mathfrak P([L]^K)$ into $\mathbb R^{K\cdot L}$ introduced in Section \ref{sec:notation-and-conventions}.
\\\\
It may be speculated whether there is a simpler description of the $DCS(L,K,L')$ than the one given through calculation of all its points $(W_1\otimes\ldots W_K)p^{(K)}$. Especially mutual information has turned out to be an important concept in various scenarios. Since the original source $p$ can, in our model, only be accessed via the channel $(\otimes_{i=1}^KW_i)$, it makes little sense to look at mutual information between the in- and the output of the system. We provide here a statement showing that the pairwise mutual informations of the output systems are also not relevant quantities in our scenario:
\begin{lemma}[Positivity of Mutual Information is not Sufficient]\label{lemma:positivity-of-MI-not-sufficient}
Consider $L=4$, and let $L'\geq2$. For every $K\in\nn$, there exist channels $W_1,\ldots W_K\in\mathcal W([L],[L'])$ such that the random variables $Y_1,\ldots,Y_K$ defined by $\mathbb P(Y_1,\ldots,Y_K=i_1,\ldots,i_K):=(\otimes_{i=1}^KW_i)p^{(K)}(i_1,\ldots,i_K)$ satisfy $I(Y_i;Y_j)>0$ for all $i\neq j\in\{1,\ldots,K\}$ and $p\in\mathfrak P_>([L])$, but $p$ cannot be inferred from $(Y_1,\ldots,Y_K)$.
\end{lemma}
\end{section}
\begin{section}{Proofs\label{sec:proofs}}
This section contains the proofs of our results, in the same order as they were stated in the previous section.
\begin{proof}[{\bf Proof of Theorem \ref{theorem:uniqueness-of-solution}}] As mentioned already, we restrict the proof to the case $K=3$. The general case is a straightforward generalization. We start by considering the bivariate cases:
\begin{align}
p^{(2)}=X_i\otimes X_j(p'^{(2)}),\qquad i\neq j.
\end{align}
These arise from the case with $K=3$ by taking the partial trace. In this case, since all the alphabets are equal, $\tr_i$ denotes the trace over the $i$-th copy of $[L]$. For example for $i=1$, $j=2$ and $k=3$ this works as follows: First we have
\begin{align}
p^{(2)}&=\sum_{i=1}^Lp(i)\delta_i\otimes\delta_i\\
&=\sum_{i=1}^Lp(i)\tr_3\delta_i\otimes\delta_i\otimes\delta_i\\
&=\tr_3\sum_{i=1}^Lp(i)\delta_i\otimes\delta_i\otimes\delta_i\\
&=\tr_3p^{(3)}
\end{align}
and, since for every $i\in[L]$ we have
\begin{align}
\tr_3 X_3\delta_i&=\sum_{j=1}^L(X_3)_{ji}=1
\end{align}
we also get
\begin{align}
(X_1\otimes X_2)p'^{(2)}&=(X_1\otimes X_2)\sum_{i=1}^Lp'(i)\delta_i\otimes\delta_i\\
&=(X_1\otimes X_2)\sum_{i=1}^Lp'(i)\tr_3\delta_i^{\otimes 2}\otimes X_3\delta_i\\
&=\tr_3(X_1\otimes X_2\otimes X_3)\sum_{i=1}^Lp'(i)\delta_i^{\otimes 3}\\
&=\tr_3(X_1\otimes X_2\otimes X_3)p'^{(3)},
\end{align}
thus we can use the assumed validity of the equation $p^{(3)}=(X_1\otimes X_2\otimes X_3)p'^{(3)}$
\begin{align}
p^{(2)}&=\tr_3p^{(3)}\\
&=\tr_3(X_1\otimes X_2\otimes X_3)p'^{(3)}\\
&=(X_1\otimes X_2)p'^{(2)},
\end{align}
as desired. The same argument works with any of the other pairs. Our goal is to show first that the validity of these $3$ pairwise equations already ensures that $X_1=X_2=X_3$. We argue as follows. First, fix $i\in[3]$ and choose any $j\neq i$ in $[3]$. Then define the matrix $\tilde X_i$ by $(\tilde X_i)_{mn}:=p'(n)(X_i)_{mn}$. Note that the columns of $\tilde X_i$ form a linearly independent set (since $X_i$ is invertible), and thus $\tilde X_i$ is invertible as well. For this argument to be valid, we do of course require our assumption $p(i)>0$ $\forall i\in[L]$ to hold true. We may now rewrite above equation slightly:
\begin{align}
p^{(2)}&=X_i\otimes X_j(p'^{(2)})\\
&=(\eins\otimes X_j)(X_i\otimes\eins)(p'^{(2)})\\
&=(\eins\otimes X_j)\sum_{m=1}^Lp'(m)(X_i\otimes\eins)\delta_m^{\otimes 2}\\
&=(\eins\otimes X_j)\sum_{m,n=1}^Lp'(m)(X_i)_{nm}\delta_n\otimes\delta_m\\
&=(\eins\otimes X_j)\sum_{m,n=1}^L(\tilde X_i)_{nm}\delta_n\otimes\delta_m\\
&=(\eins\otimes X_j)\sum_{n=1}^L\delta_n\otimes\tilde X_i^\top\delta_n\\
&=(\eins\otimes X_j)(\eins\otimes \tilde X_i^\top)\sum_{n=1}^L\delta_n^{\otimes 2}\\
&=(\eins\otimes (X_j\circ\tilde X_i^\top))\sum_{n=1}^L\delta_n^{\otimes 2}.
\end{align}
It is clear that $X_j\circ\tilde X_i^\top$ is still invertible, hence it seems convenient for the moment to have a look at the equation
\begin{align}
\sum_{i=1}^Lp(i)\delta_i^{\otimes 2}=(\eins\otimes X)\sum_{i=1}^L\delta_i^{\otimes 2},\qquad X\in Gl(L).
\end{align}
Writing this out in coordinates immediately yields
\begin{align}
p(i)&=\sum_{m=1}^L\delta(i,m)X_{im}=X_{ii}\qquad\forall i\in[L]\\
0&=\sum_{m=1}^L\delta(i,m)X_{jm}=X_{ij}\qquad\forall i\neq j\in[L].
\end{align}
Therefore, $X$ is uniquely determined through $p$ in that equation. It follows that all the $X_j\circ\tilde X_i^\top=Z_i$ for some $Z_i$, independent from the choice of $j$, and hence $X_j=Z_i\circ(\tilde X_i^\top)^{-1}$ (independent from the choice of $j$). Playing this trick two times shows that, in fact, $X_1=X_2=X_3=:X$ for some $X\in Gl_1(L)$.\\
We can now proceed to the second part of our proof.\\
Consider the equations
\begin{align}\label{eqn:fundamental-equaiton-2}
p^{(t)}=X^{\otimes t}p'^{(t)},\qquad t\in[3],
\end{align}
where $p,p'\in\mathfrak P([L])$ and $X\in Gl_1(L)$. What are possible solutions of equation (\ref{eqn:fundamental-equaiton-2}) if we require $p\in\mathfrak P_>([L])$ and $X\in Gl_1([L])$?\\
First, let us treat $p$ and $p'$ as fixed, and ask for solutions $X$. Define matrices $A,B\in Gl(L)$ by their entries $a_{ij}:=\delta(i,j)\sqrt{p(i)}$ and $b_{ij}:=\delta(i,j)\sqrt{p'(i)}$. Then we can set $t=2$ and reformulate our equation (\ref{eqn:fundamental-equaiton-2}) as
\begin{align}
\sum_{i=1}^L\delta_i^{\otimes 2}=(A^{-1}XB\otimes A^{-1}XB)\sum_{i=1}^L\delta_i^{\otimes 2}.
\end{align}
Thus a more convenient formulation of problem (\ref{eqn:fundamental-equaiton-2}) is to solve the equation
\begin{align}
\sum_{i=1}^L\delta_i^{\otimes 2}=(Y\otimes Y)\sum_{i=1}^L\delta_i^{\otimes 2}
\end{align}
for $Y$, and we can now use the same trick as before and obtain $YY^\top=\eins$. But this is equivalent to stating that $Y$ is orthogonal! Rewinding things, we know now that there exists an orthogonal matrix $Y$ such that $X=AYB^{-1}$.\\
We now use equation (\ref{eqn:fundamental-equaiton-2}) with $t=3$. We then get, using the special form of $X$ that we just obtained, the equation
\begin{align}
\sum_{i=1}^L\frac{1}{\sqrt{p(i)}}\delta_i^{\otimes 3}=Y^{\otimes 3}\sum_{i=1}^L\frac{1}{\sqrt{p'(i)}}\delta_i^{\otimes 3}.
\end{align}
Looking at specific entries, we see that the following are valid.
\begin{align}
\frac{1}{\sqrt{p(i)}}&=\sum_{j=1}^L\frac{1}{\sqrt{p'(j)}}Y_{ij}^2Y_{ij}\qquad\forall i\in[L]\\
0&=\sum_{j=1}^L\frac{1}{\sqrt{p'(j)}}Y_{ij}^2Y_{kj}\qquad\forall i\neq k\in[L].
\end{align}
However, the vectors $v_i:=\sum_{j=1}^LY_{ij}\delta_j$ ($i\in[L]$) form an orthonormal set (since $Y$ is orthogonal). Define the vectors $w_i:=\sum_{j=1}^L\frac{1}{\sqrt{p'(j)}}Y_{ij}^2\delta_j$ ($i\in[L]$), then the above can be reformulated as
\begin{align}
\frac{1}{\sqrt{p(i)}}&=\langle w_i,v_i\rangle\qquad\forall i\in[L]\\
0&=\langle w_i,v_k\rangle\qquad\forall i\neq k\in[L].
\end{align}
But this clearly implies that, for each $i\in[L]$, we have the equalities
\begin{align}
\frac{\sqrt{p'(j)}}{\sqrt{p(i)}}Y_{ij}=Y_{ij}^2\qquad\forall j\in[L].
\end{align}
Whenever a $Y_{ij}$ is zero, these are trivially true. If $Y_{ij}\neq0$, then we may divide by it and therefore obtain that, for each $i,j\in[L]$, either
\begin{align}
Y_{ij}=0\qquad\mathrm{or\ else}\qquad Y_{ij}=\sqrt{\frac{p'(j)}{p(i)}}.
\end{align}
Now we introduce sets $\mathcal I_i\subset[L]:=\{j:Y_{ij}\neq0\},\ i\in[L]$ such that the vectors $v_i$ fulfill
\begin{align}
v_i=\sum_{j\in I_i}\sqrt{\frac{p'(j)}{p(i)}}\delta_i.
\end{align}
Of course, since $Y$ is an orthonormal matrix, these vectors again form an orthonormal set. Also, it holds that each $\mathcal I_i$ satisfies $|\mathcal I_i|>0$. Hence for $i\neq l$,
\begin{align}
0=\sum_{j\in \mathcal I_i\cap\mathcal I_l}\sqrt{\frac{p'(j)}{p(i)}}\sqrt{\frac{p'(j)}{p(l)}}=\sum_{j\in \mathcal I_i\cap\mathcal I_l}p(j).
\end{align}
It follows that $\mathcal I_j\cap\mathcal I_k=\emptyset$, whenever $j\neq k$. Clearly then, since each of the sets $\mathcal I_i$ is non-empty, they all must be one-element sets. This also directly implies that, for $i\in\mathcal I_j$, it holds $Y_{ij}=1$ (meaning that $Y$ is a permutation matrix) and, additionally, that $p(i)=p'(j)$ whenever $i\in\mathcal I_j$. The desired permutation is conveniently defined through its action on $\mathfrak P([L])$:
\begin{align}
\tau(\delta_i):=\eins_{\mathcal I_i}.
\end{align}
This proves the equation \eqref{eqn:invertibility-L-L} in Theorem \ref{theorem:uniqueness-of-solution}.\\
How validity of \eqref{eqn:invertibility-L-L} implies the existence of an inverse to the map defined in \eqref{eqn:def-of-theta} can be seen as follows: Assume there are choices $W_1\ldots,W_K\in\mathcal W(([L],[L])$ and $V_1,\ldots,V_K\in\mathcal W([L],[L])$ of invertible channels and two distributions $p,p'\in\mathfrak P_>([L])$. Then our original question that we formalized in \eqref{eqn:basic-question} can be recast as follows: Can it happen that the equality
\begin{align}
\left(\bigotimes_{i=1}^KV_i\right)\sum_{i=1}^Lp(i)\delta_i^{\otimes K}=\left(\bigotimes_{i=1}^KW_i\right)\sum_{i=1}^Lp'(i)\delta_i^{\otimes K}
\end{align}
holds? Since all channels are invertible, one can equivalently ask whether the equation
\begin{align}\label{eqn:equality-of-output}
\left(\bigotimes_{i=1}^KV_i^{-1}\circ W_i\right)\sum_{i=1}^Lp'(i)\delta_i^{\otimes K}=\sum_{i=1}^Lp(i)\delta_i^{\otimes K}
\end{align}
holds. The answer to this question has already been given now and lets us conclude that there is a permutation $\tau\in S_L$ such that $V_i^{-1}\circ W_i=\tau^{-1}$ for all $i=1,\ldots,K$ and $p'=\tau(p)$. In particular, this implies that for all $i\in[K]$ we have
\begin{align}
V_i=W_i\circ\tau,
\end{align}
so that the channels differ only by one joint permutation. If one additionally assumes that $p,p'\in\mathfrak P^\downarrow([L])\cap\mathfrak P_>([L])$, then it follows that \eqref{eqn:equality-of-output} can only hold if $\tau$ is the identity map on $[L]$ and, therefore, it holds that $V_i=W_i$ for all $i\in[K]$ and $p=p'$. Thus, the function $\Theta$ is injective.\\
It remains to define $\Theta'$: Define, for $q\in\mathfrak P([L]^K)$, $\Theta'(p)$ to be the choice $p\in\mathfrak P_>([L])$ that achieves the minimum in
\begin{align}
\min_{p\in\mathfrak P_>([L])}\min_{W_1,\ldots,W_K}\|(\otimes_{i=1}^KW_i)p^{(K)}-q\|_1.
\end{align}
It is understood that the minimization is over invertible channels $W_1,\ldots,W_K$ only. Under the same assumptions on $p,W_1,\ldots,W_K$ as those made for the definition of $\Theta'$, the output $\Theta''(q)$ of the extended version $\Theta''$ is conveniently defined as the minimizer $(p,W_1,\ldots,W_K)$ in
\begin{align}
\min_{(p,W_1,\ldots,W_K)}\|(\otimes_{i=1}^KW_i)p^{(K)}-q\|_1.
\end{align}
One may replace $\|\cdot-q\|_1$ by any other norm in the above definitions, or by Kullback-Leibler divergence $D(\cdot||q)$, if additional requirements in the computation of $\Theta'$ or $\Theta''$ are required.
\end{proof}
We will now deliver the proofs for situations where the output systems are strictly smaller than the input. The basic idea is that the observing agents report events that are elements of a system with output $L'\leq L$. The question then is, whether their findings are compatible with the assumption that $L=L'$.
\begin{proof}[{\bf Proof of Theorem \ref{theorem:the-perfect-conspiracy}}]
Let the output distribution of the system be $q\in\mathfrak P([L'])^K$, and assume that $L'\leq L$ and that there are invertible channels $W_1,\ldots,W_K\in\mathcal W([L'],[L'])$ and $r\in\mathfrak P([L'])$ and $V_1,\ldots,V_K\in\mathcal W([L],[L'])$ each having full range ($\dim\ran(W_i)=\mathbb R^{L'}$) and $s\in\mathfrak P([L])$ such that both
\begin{align}
q&=(\otimes_{i=1}^KW_i)r^{(K)}\ \ \mathrm{and}\ \ q=(\otimes_{i=1}^KV_i)s^{(K)}.
\end{align}
Then, define $R:=\sum_{i=1}^{L'}r(i)^{1/2}E_{ii}$ and $S:=\sum_{i=1}^Ls(i)^{1/2}E_{ii}$ and $X_i:=R^{-1}\circ W_i^{-1}\circ V_i\circ S$. Clearly every $X_i$ is an element of $M(L,L')$. It holds, by playing the same tricks as before:
\begin{align}\label{eqn:fundamental-equations-for-degenerate-case}
X_iX_j^\top=\eins_{\mathbb C^{L'}},\qquad\forall\  i\neq j\in[K].
\end{align}
Now assume $K\geq 3$, and pick the indices $1,2,3$ as an example. Let the columns of $X_3^\top$ be $c_1,\ldots,c_{L'}$. These are linearly independent, since $X_3$ is full-range by assumption. They span the subspace $C\subset \mathbb R^{L}$ with $\dim C=L'$. From above equations (\ref{eqn:fundamental-equations-for-degenerate-case}) they further fulfill
\begin{align}
X_i c_j=e_j,\qquad i=1,2,\ j=1,\ldots,L'.
\end{align}
But that implies that $X_1c=X_2c$ for all $c\in C$. Since $\dim\ran X_i= L'$ we get $X_1=X_2$, hence especially $\ker X_1=\ker X_2$. This directly implies $\ker V_1=\ker V_2$. The same argument holds with any different choice of indices $i,j,k\in[K]$, thus the desired
\begin{align}
\ker V_1=\ldots=\ker V_K
\end{align}
follows.
\end{proof}
\begin{proof}[{\bf Proof of Theorem \ref{theorem:the-case-L'<L}}]
If $L'>L$, and $K\geq3$, we can use all our previous reasoning to do the following steps:
First, for two $DCA(L',K,L)$ systems $(p,W_1,\ldots,W_K)$ and $(p^\backprime,V_1,\ldots,V_K)$ we get
\begin{align}
W_i^{-1}\circ V_i=W_j^{-1}\circ V_j\qquad \forall i,j\in[K].
\end{align}
With $X:=\sqrt{p^{-1}}\circ W_1^{-1}\circ V_1\circ\sqrt{p^{\backprime}}$ we then get, as before, that $X=\tau$ for some permutation $\tau$ on $[L']$, and it follows that the two systems are equivalent up to permutations.
\end{proof}
\begin{proof}[{\bf Proof of Theorem \ref{theorem:inversification}}] We will first and in greater depth consider the case $L'=2$. In a second step, we generalize our proof to arbitrary $L'$.\\
Let $\{\gamma_i\}_{i=1}^{L}\subset\mathbb R$ be such that
\begin{align}
0=\sum_{i=1}^{L}\gamma_iW^{\otimes K}(\delta_i^{\otimes K}).
\end{align}
We can assume that we have $W(\delta_i)=\lambda_i\delta_1+\lambda_i'\delta_2$ for an appropriate set of $\{\lambda_i\}_{i=1}^L\subset[0,1]$ and with the convention $\lambda_i':=1-\lambda_i$ for all $i\in[L]$. We also know that the set $\{\delta_1^{\otimes t}\otimes \delta_2^{\otimes K-t}\}_{t=0}^K$ is linearly independent. Thus, we get for every $t\in\{0,\ldots,K\}$:
\begin{align}
0&=\sum_{i=1}^L\gamma_i\lambda_i^t(\lambda_i')^{K-t}.
\end{align}
This set of equations can easily be seen to be equivalent to
\begin{align}\label{eqn:Bernstein-equation}
0&=\sum_{i=1}^K\gamma_i\frac{1}{{K\choose t}}B_{t,K}(\lambda_i)\qquad\forall t=0,\ldots,K,
\end{align}
where $B_{t,K}$ are the Bernstein-polynomials \cite{bernstein} defined as $B_{t,K}(x):={K\choose t}x^t(1-x)^{K-t}$ ($x\in\mathbb R$). It is known\footnote{This fact was first brought to our attention through the "On-Line Geometric Modeling Notes'' of Kenneth Joy at UC Davis.} \cite{farouki} that these polynomials span the space $Pl(K)$ of all polynomials of degree no more than $K$. We can therefore reformulate (\ref{eqn:Bernstein-equation}) as
\begin{align}
0=\sum_{i=1}^L\gamma_iP(\lambda_i)\qquad \forall P\in Pl(K).
\end{align}
Given a choice of the $\gamma_i$, we can always find a polynomial $P$ satisfying $P(\lambda_i)=\gamma_i$, as long as $K-1\geq L$. Then, we are left with the equation
\begin{align}
0=\sum_{i=1}^L\gamma_i^2,
\end{align}
which clearly implies $\gamma_1=\ldots=\gamma_L=0$, hence $W^{\otimes K}$ is invertible.
\\\\
In the more general setting where $L>L'$ and with $W$ having full-range image we know there exist points $\{\mathbf a_i\}_{i=1}^{L}\subset\mathbb R^{L'-1}$ such that for the output distributions $W(\delta_i)$ it holds, with $a_{i,0}:=1-\sum_{j=1}^{L'-1}a_{i,j}$: $W(\delta_i)=\sum_{j=1}^{L'-1}a_{i,j}\delta_j+a_{i,0}\delta_{L'}$.\\
Then as before, we can rewrite the statement
\begin{align}
0=\sum_{i=1}^L\gamma_iW^{\otimes K}(\delta_i^{\otimes K})
\end{align}
as
\begin{align}
0=\sum_{i=1}^{L}\gamma_iB_{f,K}(\mathbf a_i)\qquad\forall\ f,
\end{align}
where $B_{f,K}$ are the multivariate Bernstein polynomials and $f:[L]\to\nn$ are so-called frequencies and satisfy $f\geq0$ as well as $\sum_{i=1}^Lf(i)=K$.
But for a fixed choice of the $\gamma_i$, this means that the statement
\begin{align}
0=\sum_i\gamma_iP(\mathbf a_i)
\end{align}
has to hold \emph{for all} linear combinations $P$ of the polynomials $B_{f,K}$. It is known \cite{jetter-stoeckler} that these span the space of all polynomials of degree less than or equal to $K$. Hence, above equality carries over to all polynomials $P$ in $L$ variables of degree less than or equal to $K$. In the general case on can (based on the theory of Kergin-interpolation \cite{kergin,sauer-xu}) show that we can always find a polynomial $P_\gamma$ satisfying $P_\gamma(\mathbf a_i)=\gamma_i$, $i=1,\ldots,L$, if $K\geq L(L'-1)$ holds.
\end{proof}
We will now come to the last remaining one of our proofs.
\begin{proof}[{\bf Proof of Lemma \ref{lemma:positivity-of-MI-not-sufficient}}]
Just let each $W_1,\ldots,W_K=W$ where $W$ is defined by
\begin{align}
W(\delta_1)=\delta_1,\ \ W(\delta_2)=\delta_2,\ \ W(\delta_3)=W(\delta_4)=r,
\end{align}
where $r\in\mathfrak P([L])$ is arbitrary. Then, it is impossible to tell the values $p(4)$ and $p(3)$ of the sought-after $p$: think of the sets
\begin{align}
P_{ab}:=\{p:p(1)=a,\ p(2)=b\}.
\end{align}
Every of these distributions gets mapped to the same distribution
\begin{align}
a\cdot\delta_1^{\otimes K}+b\cdot\delta_2^{\otimes K}+(1-a-b)r^{\otimes K}\otimes r^{\otimes K}
\end{align} by application of above channels. Thus, we can \emph{never} hope to get an invertibility criterion just from looking at the pairwise mutual informations!
\end{proof}
\end{section}
\begin{section}{Appendix\label{sec:appendix}}
We now briefly touch upon the topics hypothesis testing and statistical inference.
\begin{subsection}{Hypothesis testing for dependent component systems}
In this section, we present some facts on hypothesis testing in order to connect this presentation to it.
Given an arbitrary $DCS$ $\mathcal S$ with parameters $L,K,L'$, what we receive at the output during $n\in\nn$ observations is a string $y^n\in([L']^n)$. Let the number of times each symbol $(y_1,\ldots,y_K)\in[L']^K$ appears in $y^n=((y_{1,1},\ldots,y_{K,1}),\ldots,(y_{1,n},\ldots,y_{K,n}))$ be $N(y_1,\ldots,y_K|y^n)$. Due to the memoryless nature of the system, every permutation of that string is equally likely to be the output of the system. Hence, the whole set $\{\tau y^n:\tau\in S_n\}$ (where $S_n$ is the set of permutations on $n\in\nn$ symbols and $(\tau y^n)_{i}:=(y_{1,\tau^{-1}(i)},\ldots,y_{K,\tau^{-1}(i)})$) will get mapped to the same estimate $\hat q=\hat q(y^n)$. We will henceforth, for nonnegative functions $f:[L']^K\to\mathbb N$, use the abbreviation
\begin{align}
T_{f}:=\{y^n:N(\cdot|y^n)=f\}.
\end{align}
Note that $T_{N(\cdot|y^n)}=\{\tau y^n:\tau\in S_n\}$ holds. What is the best (asymptotic) estimate, given $y^n$? This question is answered as follows: We search for
\begin{align}
\hat q:=\arg\max\{q^{\otimes n}(y^n):q\in\mathfrak P([L'])^K)\}.
\end{align}
According to \cite[Proof of Lemma 2.3]{csiszar-koerner}, the solution to this optimization problem is given by the maximum-likelihood estimate $\hat q=\tfrac{1}{n}N(\cdot|y^n)$. Moreover, the probability that this estimate $\hat q$ satisfies $D(\hat q\|q)>\eps$ for some $\eps>0$ is bounded by
\begin{align}
q^{\otimes n}(\{y^n:D(\tfrac{1}{n} N(\cdot|y^n)\|q)>\eps\})&\leq poly(n)2^{-n\cdot\eps},
\end{align}
where $poly(n)$ denotes a polynomial in $n$ that depends on $L$ as well. Obviously, this upper bound goes to zero exponentially fast in $n$. One may choose $\eps$ depending on $n$ by setting e.g. $\eps_n:=1/\sqrt{n}$, and this delivers a hypothesis test that succeeds with probability going to one as $n$ tends to infinity.\\
Adopted to our scenario, it will identify the output distribution $q$ of any $DCS$ with arbitrary precision. It remains to be proven that small errors in the estimate remain bounded when inverting the system.
Also, the question of optimality of this choice of test remains.\\
In hypothesis testing scenarios as described for example in \cite{blahut}, one typically considers binary hypotheses. That is, one assumes that the true state of the system is given by either of two distribution $r,s$. In the scenario described in this work however, we are faced with an additional subtlety: The channels that map the system outputs to the observer are unknown as well.\\
Thus a rigorous problem formulation in our scenario includes the adoption of additional hypotheses on the channels, may these be of the nature 'it is either $\otimes_i W_i$ or $\otimes_i V_i$' or 'the channels are drawn at random according to some distribution on the set of channels'. A worst-case assumption would be that every of the channels $\otimes_iW_i$ is possible. In that case we encounter an additional problem: The quantity
\begin{align}\label{eqn:our-hypothesis-testing-lemma}
\inf_{\mathfrak W,\mathfrak V\in\mathcal W_{\leftrightarrow}(L,K)}D(\mathfrak Wr^{(K)}\|\mathfrak Vs^{(K)}),
\end{align}
where $\mathcal W_{\leftrightarrow}(L,K)\subset\mathcal W([L],[L])^{\otimes K}$ is the subset of invertible channels of the form $\mathfrak W=W_1\otimes\ldots\otimes W_K$ is simply equal to zero. This follows from the fact that in every vicinity of a non-invertible channel there is an invertible channel, too.
\end{subsection}
\begin{subsection}{Connection to statistical inference\label{sec:simpson-paradox}}
In this subsection we briefly connect our findings to the area of statistical inference. Let for simplicity $\bA,\bB,\bC$ be binary alphabets. We let $\bC$ be the input of a $DCS$ and $\bA$, $\bB$ the output systems. Let $q\in\mathfrak P(\bA,\bB)$ be the output of the $DCS$. Let the overall distribution be $s\in\mathfrak P(\bA\times\bB\times\bC)$. Due to the special structure of our system we have
\begin{align}\label{eqn:our-special-strucutre-in-simpsons-paradox}
s_{A|BC}=s_{A|C}.
\end{align}
It is also clear by construction that the events happening on $\bC$ are a common cause for those on $\bA$ and $\bB$. As explained in \cite{frosini,reichenbach} such systems go under the name 'conjunctive fork'. They are to be distinguished from systems which are included under the name 'Simpson's paradox'. While in our case we have two channels going (strictly speaking) in parallel from $\bC\times\bC$ to $\bA\times\bB$, a system on which Simpson's paradox (wrongly inferring that some statistical event is causal for another statistical event) can occur would for example have to have a Markov structure $\bC\to\bA\to\bB$.
\end{subsection}
\end{section}
\begin{section}{Open problems}
Stability: Once we have estimated the output distribution on $[L']^K$ we would like to invert it in order to know $p$. Then, if a small mistake in the estimation scheme would lead to dramatically different results for $p$, we would rightfully see this as a drastic drawback of the method. This issue deserves further attention.\\
Hypothesis testing: It would be desirable to compare different hypothesis testing scenarios and derive optimal tests for them. It remains to be seen whether this can lead to the derivation of new and meaningful information measures. For more information, see the appendix.\\
Activation: We left open the question of a general 'activation' effect of invertibility. It would be interesting to know under what conditions a general set of $K$ channels $W_1,\ldots,W_K\in\mathcal W([L],[L'])$ becomes invertible as a map $\otimes _{i=1}^KW_i$ from $\mathfrak P^{(K)}([L])$ to $\mathfrak P([L']^K)$, for general $L$ and $L'$. This includes a more detailed study of the projective behaviour that was only touched upon in Remark \ref{remark:onL>L'}.\\
Multivariate polynomials: A detailed investigation of this connection is postponed to future work.\\
Another possible route for future research is the connection of our findings to the very structure of three-dimensional space as experienced by us every day, as has been done e.g. in \cite{mueller-masanes} for qubits.\\
At last we have to mention that, of course, an analysis of infinite-dimensional $DCS$s, extensions to quantum mechanics and an investigation of $DCS$ with time varying channels or distributions offer the potential of finding results that are interesting in their own right.
\end{section}

\begin{IEEEbiographynophoto}{Janis N\"otzel} received the Dipl. Phys. degree in physics from the Technische Universit\"at Berlin, Germany, in 2007 and the PhD degree from Technische Universit\"at M\"unchen, Germany, in 2012. From 2008 to 2010 he was a research assistant at the Technische Universit\"at Berlin, Germany, and from 2011 until 2015 at Technische Universit\"at M\"unchen. From July 2015 to August 2016 he was a DFG research fellow at Universitat Aut\`{o}noma de Barcelona, Spain. He is now with the Technische Universit\"at Dresden, Communications Laboratory, 01069 Dresden, Germany.
\end{IEEEbiographynophoto}
\begin{IEEEbiographynophoto}{Walter Swetly} studied Electrical Engineering, Logic, Philosophy and Theoretical Linguistics at Technische Universit\"at München and Ludwig-Maximilians-Universit\"at M\"nchen. He received his PhD degree from LMU M\"nchen in Logic in 2010.
From 2010 - 2012 he was a member of the Chair for Logic and Philosophy of Science at LMU M\"unchen and from 2012 - 2014 at the Chair for Data Processing at Technische Universit\"at M\"unchen. Since then he has worked in various positions in the industry.
\end{IEEEbiographynophoto}
\end{document}